\documentclass[sigconf]{acmart}

\AtBeginDocument{%
  \providecommand\BibTeX{{%
    \normalfont B\kern-0.5em{\scshape i\kern-0.25em b}\kern-0.8em\TeX}}}

\copyrightyear{2022}
\acmYear{2022}
\setcopyright{cc}
\setcctype{by}
\acmConference[ Asian HCI Symposium'22]{ Asian HCI Symposium'22}{April 29-May 5, 2022}{New Orleans, LA, USA}
\acmBooktitle{ Asian HCI Symposium'22 ( Asian HCI Symposium'22), April 29-May 5, 2022, New Orleans, LA, USA}
\acmDOI{10.1145/3516492.3558788}
\acmISBN{978-1-4503-9250-1/22/04}

\usepackage{comment}

\begin{document}

\title[Physical and AR-based Playful Activities for training ASHAs in India]{Physical and Augmented Reality based Playful Activities for Refresher Training of ASHA Workers in India}

\author{Arka Majhi}
\email{arka.majhi@iitb.ac.in}
\affiliation{
  \institution{Indian Institute of Technology Bombay}
  \streetaddress{Powai}
  \city{Mumbai}
  \state{Maharashtra}
  \country{India}
  \postcode{400076}
}
\orcid{https://orcid.org/0000-0002-5057-1878}

\author{Satish B. Agnihotri}
\email{sbagnihotri@iitb.ac.in}
\affiliation{
  \institution{Indian Institute of Technology Bombay}
  \streetaddress{Powai}
  \city{Mumbai}
  \state{Maharashtra}
  \country{India}
  \postcode{400076}
}
\orcid{https://orcid.org/0000-0002-0703-3185}

\author{Aparajita Mondal}
\email{aparajita.mondal@iitb.ac.in}
\affiliation{
  \institution{Indian Institute of Technology Bombay}
  \streetaddress{Powai}
  \city{Mumbai}
  \state{Maharashtra}
  \country{India}
  \postcode{400076}
}
\orcid{https://orcid.org/0000-0003-4609-2249}

\renewcommand{\shortauthors}{Arka Majhi, Satish B. Agnihotri and Aparajita Mondal}

\begin{abstract}
Recent health surveys in India highlight the alarming child malnutrition levels and lower rates of complete child immunization in many parts of India. Previous researches report that the conventional training pedagogy of the CHWs (Community Healthcare Workers) or the ASHAs (Accredited Social Health Activists) in India is ineffective in enhancing their capacity. Considering that the CHWs are getting equipped with smartphones, it calls for a rethinking of their training pedagogy using the ICT approach. Two refresher training tools were developed to make learning the child immunization schedule more exciting and conceptually engaging for ASHAs. The physical and AR (Augmented Reality) versions of designed card games were compared for effectiveness and knowledge retention, pre, and post-intervention through questionnaire tests conducted immediately before and after playing multiple sessions. The AR-based play was found to be better in learning and knowledge retention with more engagement, mainly due to its interactive and intuitive nature of play.
\end{abstract}



\maketitle

\section{Introduction}

 Research on the training and education of CHWs (Community Healthcare workers) is scarce. Few academic researchers have attempted to train and educate CHWs using technology solutions like ICT, community radio, and multimedia as a medium through feature phones and smartphones \citep{Yadav2017,Yadav2019,Yadav2019a,Kumar2013,Kumar2015,Kumar2015a,Javaid2017}. In this research, we designed, developed, and field-tested a refresher training game in both physical and digital versions for CHWs of West Bengal, India. Our study has been conducted in multiple phases, including conceptualizing a training game, getting the content of the game vetted, game design, game development, field testing, comparative evaluation, and pilot testing in real-world settings.
 
 The aim of this study is to make the immunization schedule more exciting and conceptually engaging by utilizing the features of card games. This refresher training game will train CHWs (Community Healthcare Workers) or ASHA (Accredited Social Health Activists) workers on immunization schedules and the outcomes if the immunizations are missed. Trainers of ASHA workers could also use this tool in addition to their conventional teaching techniques to improve retention of knowledge of ASHAs.

This research contributes to the field of HCI, HCI4D, and healthcare research in the Indian context. Our contribution to HCI and Training Education is mainly through deploying and conducting a longitudinal field study of the digital and physical card game, comparing them, and discussing the effects of player in-game behavior and activities, which results in fun, engagement, knowledge gain, and knowledge retention. After offline multimedia videos for feature phones \citep{Shah2017}, instructional and communicative illustrations \citep{Tulaskar2020} and quiz-based training app \citep{Majhi2021}, this research focuses on comparing physical and digital AR card games for training CHWs which is a natural extension to the previous studies conducted by Indian researchers over the last few years on making games and playful activities. Our research contributes to and builds on previous research studies. HCI4D community has been growing interest in CHWs, who are emergent smartphone users. The focus is on designing holistic learning solutions for them, focusing on the first 1000 days of development encompassing maternal (pregnancy, ante, and post-natal) and child (0-2 yrs) healthcare.

\section{Background}

Through immunization, 2-3 million lives are saved each year from diseases that vaccines could prevent \citep{Andre2008}. Across the world, about 19.4 million children with age less than 12 months missed the necessary vaccination, with 11.7 million from low-and-middle-income countries (LMICs) and 2.6 million of them being Indian \citep{WHO2020}. In India children aged 12-23 months receiving all essential vaccinations have increased from 44\% in 2005-06 to 62\% in 2015-16 \citep{NFHS3IndiaFactsheet}\citep{NFHS4IndiaFactsheet}. 47\% of children belonging to the lowest wealth quantile households in India are not fully immunized, and any health insurance does not cover 71\% of families \citep{NFHS4IndiaFactsheet}. As a result, they end up spending significant amounts on healthcare, pushing almost 32-39 million people annually to dip below the poverty line \citep{vanDoorslaer2006, Garg2009}. With the high penetration of smartphones not only in urban but also in rural and tribal India, mHealth technologies could help in improving health outcomes and decrease health inequalities \citep{Peek2017}.

Public healthcare systems worldwide are experiencing a lack of CHWs or skilled healthcare professionals, particularly in LMICs, with an anticipated 80 million shortage by 2030 \citep{WHO2020}. It is still difficult for LMICs to educate or train CHWs to ensure the health and well-being of the most vulnerable communities \citep{WHO2020}. A growing trend in designing technology is assisting CHWs in their job by supporting health education dissemination \citep{Vashistha2016,Kumar2015}, generating on-demand reviews of household visits \citep{DeRenzi2017}, collecting input from care recipients, and creating opportunities for training and learning \citep{Muke2020,Yadav2017,Yadav2019}. It is a challenge in developing countries like India, where a majority of CHWs are 'low-literate' users by definition \citep{Ahmed2013}. But interestingly, most of the CHWs are not digitally illiterate. They can use smartphones for some of their basic tasks and fall somewhere in the category of 'Basic User, Navigator, Text Inputter, Saver and Account Holder' as per the adoption of technology and user-usage model \citep{Devanuj2013,Doke2015}. Finance Minister of India, during the Union Budget 2020-21, announced that over 0.6 million Anganwadi workers, a cadre of CHWs in India, were smartphone equipped in 2020 for uploading the nutritional status of more than 100 million households. India has seen a constant rise in CHWs' smartphone ownership. It brings an opportunity to improve healthcare access and engagement with the beneficiaries in the community through mHealth approaches \citep{Bassi2018,Madanian2019}.

Previous researchers suggested ways to design effective methods of disseminating health education \citep{Kumar2015} through community radio, SMS/ text messages, and short videos \citep{Ramachandran2010}. Watching video clips on mother and child care encouraged CHWs in managing pregnancies better \citep{Kumar2015}. Shah et al. conducted an incentive-based gamification study \citep{Shah2017}. Videos related to mother and child healthcare were loaded on the phones of CHWs. They were later quizzed on the content they watched and awarded an incentive of cellular talk time. Tika Vaani \citep{Perez2020} an IVR-based solution, improved immunization access through timely notifications and alerts. With advancing technology, the conventional method of classroom teaching for CHWs in training centres has evolved towards interactive teaching modes like projected screens and blended learning platforms like the e-Incremental Learning Approach portal (e-ILA). These approaches continue to be trainer-led, with CHWs witnessing the trainer's activities and the resulting occurrences. As a result, maintaining the attention of the batch of CHWs during the training period becomes challenging. The quality of training could improve by implementing Activity Based Learning (ABL), which necessitates the creation of relevant physical activity and physical kits like tokens to illustrate a concept in depth and render better knowledge retention. While physical  tokens could be used to develop tabletop games, training through personal portable devices like android smartphones and tablets can happen through digital applications. In this space, apps using Augmented Reality (AR) could be considered an effective approach towards creating pedagogy for constructive learning through trials and actions. AR is an emerging technology that allows learner-centered, Activity Based Learning \citep{Dunleavy2009,Wu2013} with its ability of real-time superimposition of computer-generated virtual visuals over the actual environment \citep{Azuma1997}. Beyond being interactive, AR allows for the relatively seamless integration of digital 3D models into the actual environment, resulting in a more immersive and contextual learning experience for learners \citep{Billinghurst2012,Santos2014}. Augmented Reality Learning Experiences (ARLEs) have been defined as "the learning experience facilitated by Augmented Reality technology" \citep{Santos2014}. There have been researches that highlight learners' positivity towards creating knowledge and patterns of behavior while interacting through AR learning situations cooperatively \citep{Sarkar2020}.

\section{Method}
\subsection{Assessment of Prior Knowledge}
A Focused Group Discussion (FGD) was conducted among a group of young CHWs at Falta village in the South 24 Parganas district of West Bengal. There were 14 CHWs in total, and 11 of them were under 30 years of age group. 11 of them completed 12 years of schooling, while 3 graduated. Most of them had 5 years of experience being a CHW. We found that most CHWs lacked complete knowledge of the immunization schedule, growth monitoring chart, and other aspects of the Mother and Child Protection card (MCP card). MCP card is a key step in ensuring that a child receives all of the required immunizations on time. At the time of conducting the NFHS-4 survey, vaccination cards were available for just 63\% of children aged 12-23 months \citep{NFHS4IndiaFactsheet}. A study found that the issuing of MCP cards to pregnant mothers was strongly associated with partial immunization in children \citep{Kizhatil2019}. Thus, we chose the child immunization schedule from the MCP card as content for developing playful activities for providing refresher training for CHWs. The schedule tabulates the vaccines prescribed for a child at different ages.

\subsection{Conditions and Mechanics chosen for playful learning activity}

During FGD, the authors initiated a brainstorming session with the CHWs on ways to design playful activities. The core idea generated was that the information on vaccines has to be displayed in blocks in front of the players, from which the player has to make a decision based on the information and the scenario represented. Possibilities of play through physical cards as play tokens were explored. The primary reasons for choosing physical cards were their cost-effectiveness and handling ease. Multiplayer research related to playing with physical cards \citep{Bochennek2007} had shown significant learning advantages through engagement and collaboration.

\subsection{Designing the playing card}

The cards were designed with a focus on combining written statements and illustrations for information delivery. So we designed playing cards with graphics, text titles, and brief descriptions over them. Images are more effective in conveying information in a straightforward and attractive way than text, and people remember them easily. Images reduce language barriers and become more self-explanatory.

In the FGD authors conducted a participatory design (PD) exercise with the CHWs. Royalty-free illustrations from different web repositories were printed and displayed in front of the CHWs. They selected the most appropriate image which resembles a particular medicine or vaccine, or a clinical condition. The finally chosen illustration for the cards was the ones that were selected by the majority. Some CHWs sketched rough illustrations of their own to suggest better graphic options for cards. The co-design exercise resulted in illustrations that were uniquely suited for the context. Bengali, a local Indian language spoken by the people from West Bengal, was chosen as the primary language for the text content of the cards, as we planned to conduct field studies in rural and sub-urban regions of West Bengal.

\begin{figure}
\centering
\begin{tabular}{ccc}

\includegraphics[width=0.3\linewidth]{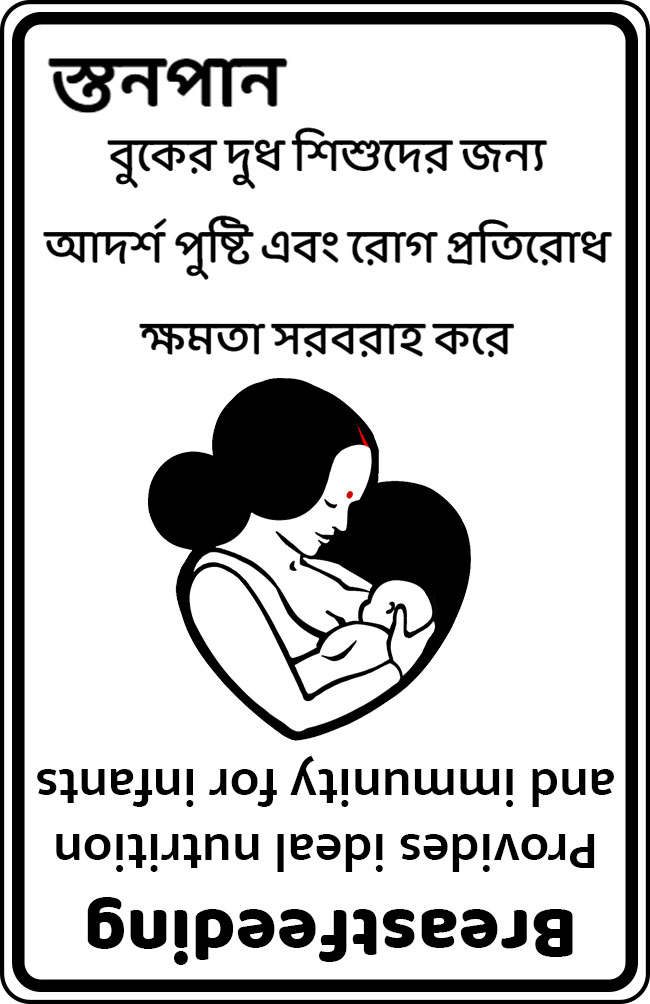}
&
\includegraphics[width=0.3\linewidth]{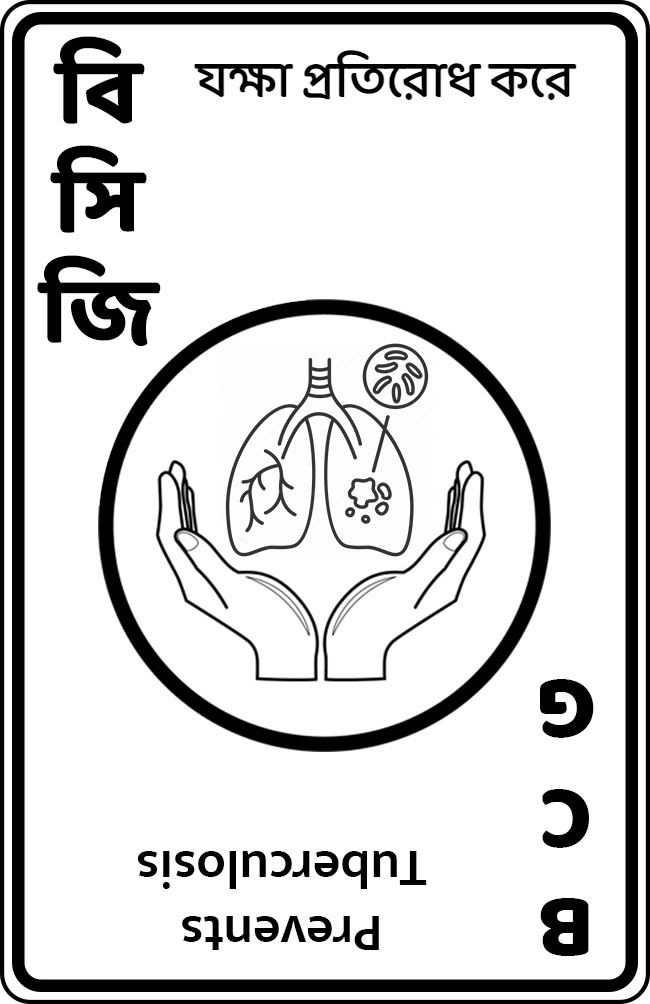}
&
\includegraphics[width=0.3\linewidth]{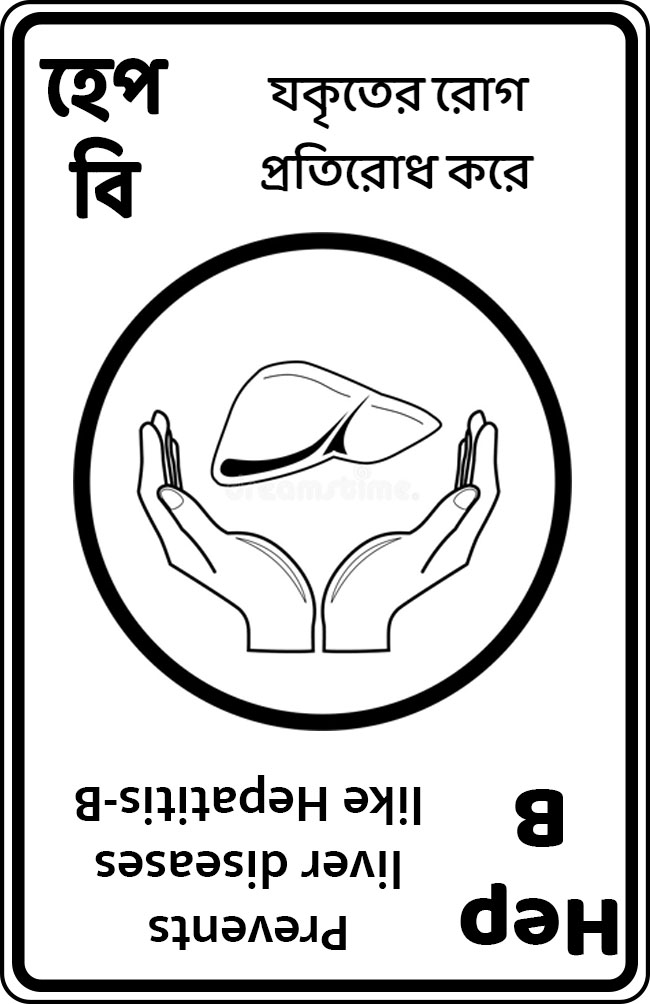}
\\
\includegraphics[width=0.3\linewidth]{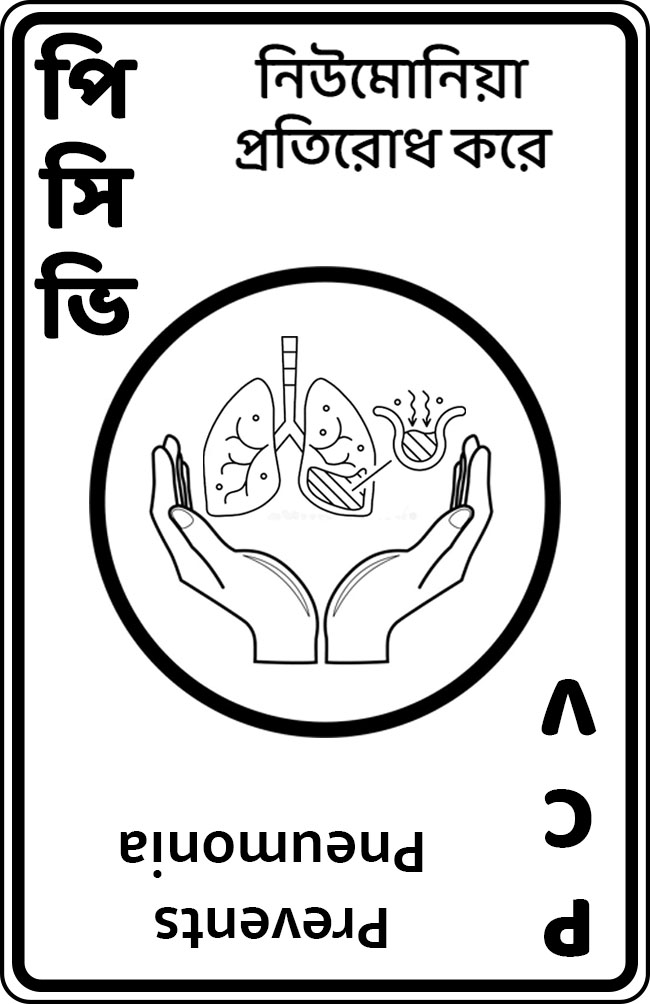}
&
\includegraphics[width=0.3\linewidth]{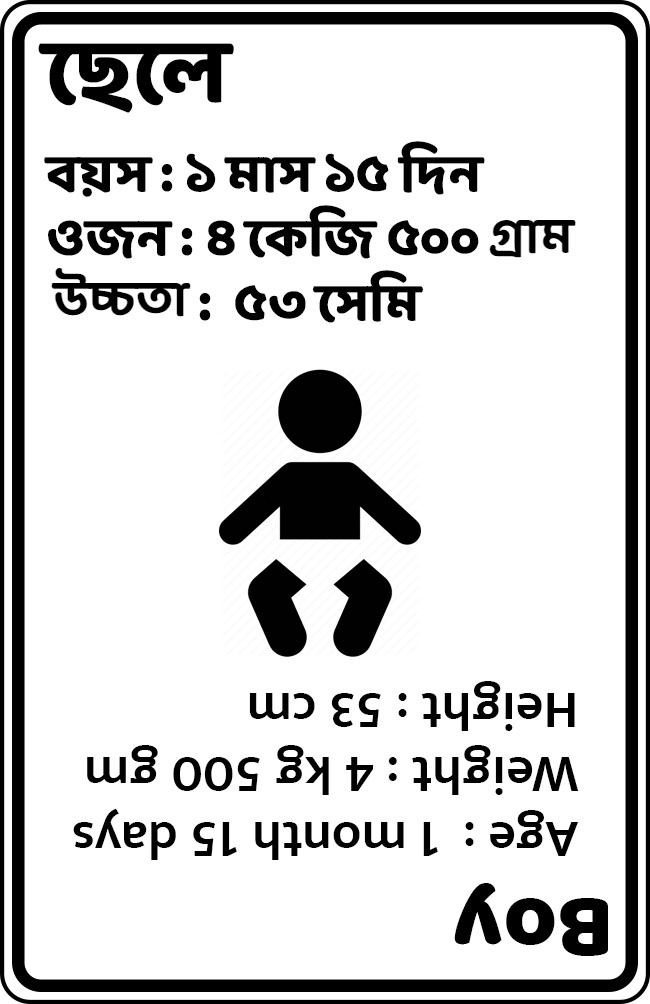}
&
\includegraphics[width=0.3\linewidth]{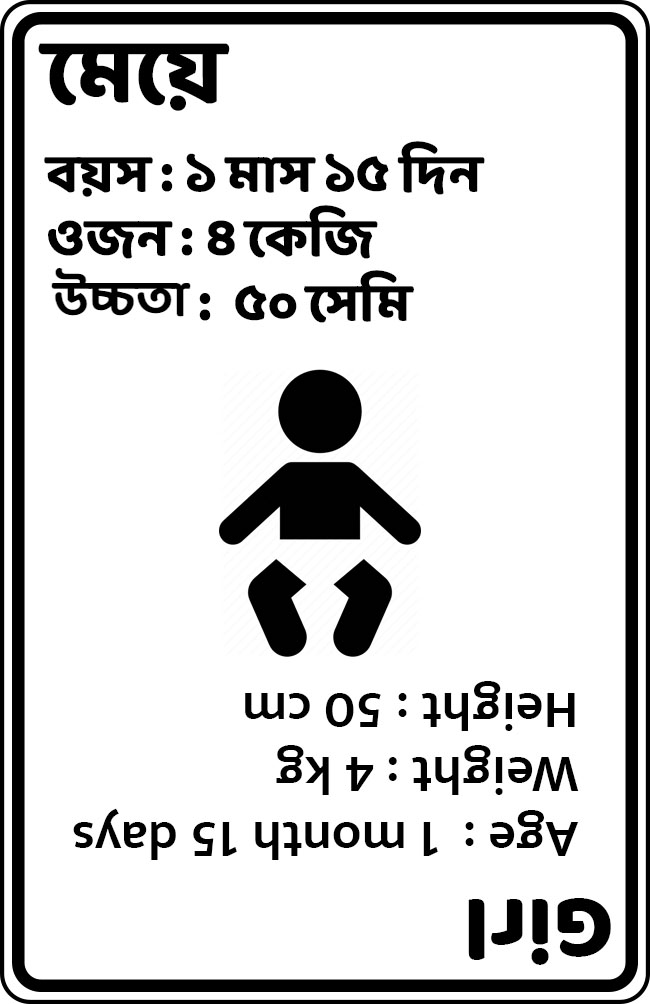}
\end{tabular}
\caption{Sample cards from the deck}
\Description{Sample of 6 cards from the deck}
\label{fig:card-deck}
\end{figure}

The card deck consisted of 20 different types of cards. There were 10 cards on immunization, 5 cards on medicines and supplements for children, 3 cards on medications and immunization for pregnant mothers, and 2 unique sets of cards for boy and girl children of varying ages, heights, and weights. The cards were of the size of a standard playing card or credit card (5.5 X 8.5 cm), made of card paper, and had printed information and graphics on them, as shown in Figure \ref{fig:card-deck}.

\subsection{Framing Challenges and Rules of the Physical Card Activity}

The primary idea of the play activity was to form sets of cards for each child comprising a child card with all vaccines which are to be given in the age group the child is in. The same goes for the pregnant mother card set.
\begin{itemize}
    \item The cards were shuffled and distributed equally among players. The players, after having a look at their cards, had to put a bet on the number of children they could immunize. The numbers were noted down. Then the child cards were distributed equally among players.
    \item When a player pulled a card from the deck, she could either get a vaccine or a medicine card. She may drop the pulled-up card if she didn't require it or thought that the odds that it might form a combination were too less. But if the player planned to retain it thinking that it might be used later, then the player would require to drop another card out of the already held ones.
    \item For showing an immunized child, the players need to put up a combination of the right set of vaccine cards corresponding to the correct child's age and read aloud the specific case.
    \item After a player has shown an immunized child; the other players can challenge the combination. The challenger has to tell the right combination or the logical combination of cards. All other players verify the contested knowledge from their own knowledge or refer to the rule book. If the combination is found wrong, the player has to pick up all the cards presented by her. As a penalty, the challenger player will get a chance to pose an adverse child health condition to the challenged player. The challenged player has to produce the medicine card. If the challenged player produces the right one, she doesn't attract a penalty, else the challenger player picks up a random card from the challenged player's hand and discards it into the pile.
    \item At the end of a play round, if the players had managed to score by showing the number of immunized children (n), they promised at the start of the game then they would get 10 times the number of immunized children (10 X n) points with an addition of 1 point as a bonus if they had produced more combinations. But if they fail to do so, then (10 X n) points are deducted from the player's total score as a penalty.
\end{itemize}
The playful activity could be played by 2–5 players and usually took about 10–15 minutes to finish each round. Though shuffling the deck brings 'luck' and 'chance' elements to the play, usually, after five rounds, a player was expected to go through all the possible combinations of sets of cards or challenges.

\subsection{Framing Challenges and Rules for the AR-based Play Activity: Tikakaran-AR}

The CHWs can either play as a single player or in multiplayer mode (depending on the availability of CHWs) with one or more silos which include pregnancy, immunization, and IYCF (Infant and Young Child Feeding). 

\begin{figure}
  \includegraphics[width=\linewidth]{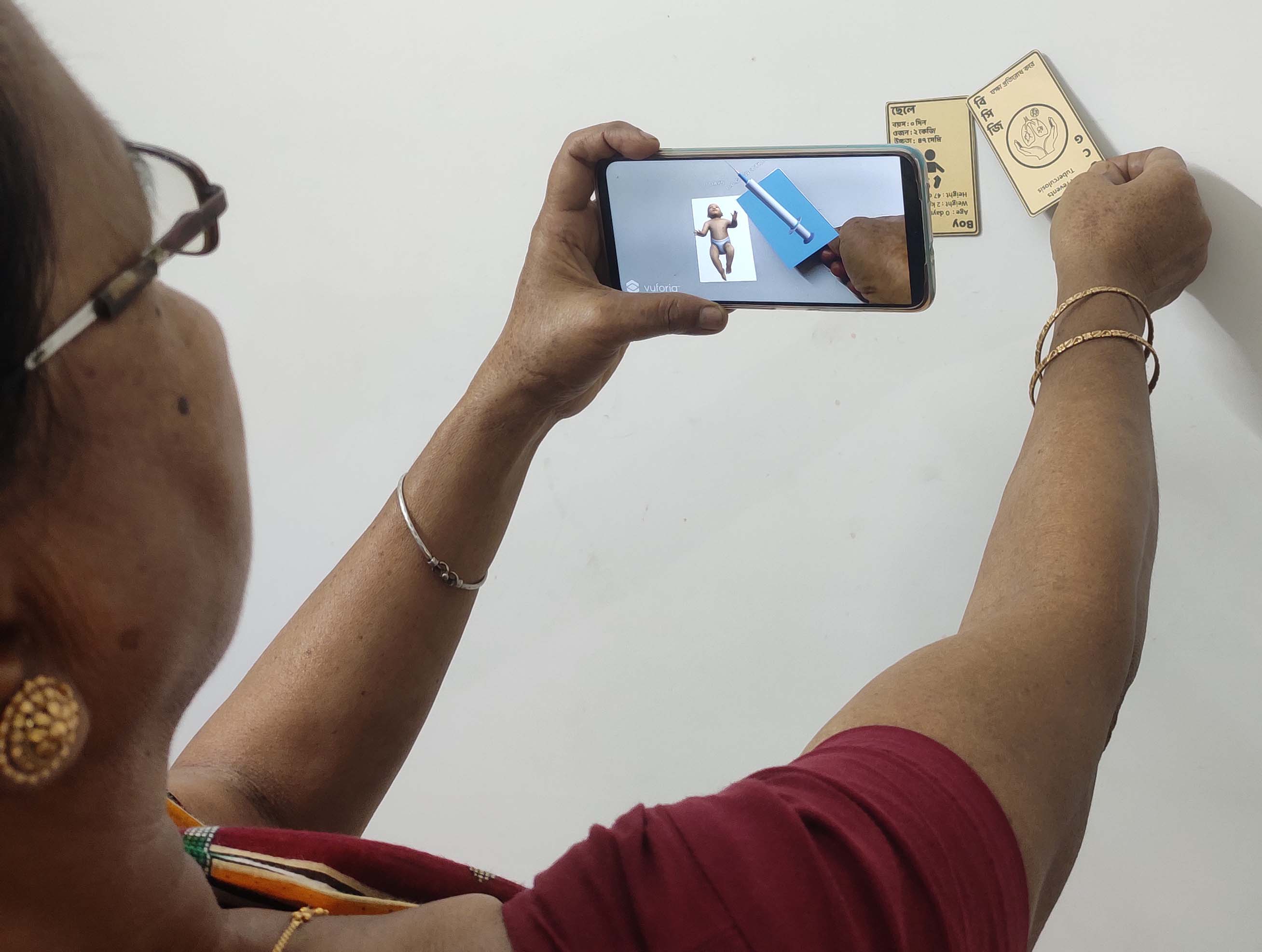}
  \includegraphics[width=\linewidth]{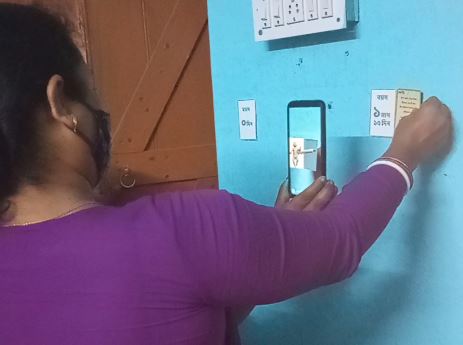}
  \caption{ASHAs play-testing the Tikakaran-AR play app}
  \Description{Top Image : An ASHA worker is holding the phone with her left hand and scanning the card she is holding with her right hand. A child card sticker is stuck on the wall.
  Bottom Image : An ASHA worker holds her smartphone in her left hand and scans the child sticker on the wall. With her right hand, she brings the BCG vaccine card beside the child card. The AR 3D model of the child shows up and is heard laughing.}
  \label{fig:teaser}
\end{figure}

\begin{itemize}
    \item Before starting the play round, the CHWs need to get the printed stickers/cards of the child and mother from the kit. They would paste these 10-20 stickers on 4 walls of a room at a height suitable for scanning through smartphones. The players will be carrying a kit with all immunization and medical supplements in the form of playing cards to be administered to the children and pregnant mothers.
    
    \item If CHWs are playing single-player, the aim of the playful activity is to immunize all children and provide care to pregnant mothers in the least time possible with the least errors as per the schedule. To initiate the play, players would hold their smartphone in one hand, start the AR play app, and scan any one sticker on the wall of their choice. On the other hand, they will browse for specific immunization cards from their bag and scan them, as shown in Figure \ref{fig:teaser}. 
    
    \item For example, if they scan a sticker with a child of 0 months, a 3D model of a child appears layered on top of the card, with age and anthropometric information over it. After seeing this, the CHW is expected to hold the phone in one hand and, with the other hand, open her bag and pull the specific cards which are appropriate for that condition (BCG, HepB, OPV, and Colostrum Feeding). If the right set of cards is placed beside the child card and scanned one by one, the points are increased with auditory feedback from a laughing child. But, if the combination goes wrong, by placing a wrong card, the child starts crying, and the points decrease, signifying a wrong choice made.
    
    \item All wrong choices are scored negatively equally. The choices which are left out in each turn also attract negative scoring. The player would proceed to the next child card and repeat it until all the children were immunized. This process of immunizing all the children counts as a round of play. 
    
    \item After one play round, the next play round starts with switching one player from each team. All rules and conditions remain the same. But, the children move to the next immunization period. For example children of \(1\frac{1}{2}\) months in round 1 turns to \(2\frac{1}{2}\) months and so on in round 2.
    
    \item Players are expected to play 4 rounds back to back or in a loop to complete a session. Some immunizations can also be carried out next month or the next round in play.
    
    \item In multiplayer mode also, the aim of the play is the same as that of a single player. But, in multi-player mode, 4 players will play in teams of 2. The resources or physical play cards will be randomized and divided into 4 decks and put in the bags of 4 players.
    
    \item During the play, if players require any other cards to complete the stage, they can exchange one card hand to hand with their partner in the group (within the group) or from players of another team (between groups). The exchanges can happen as many times throughout the play round.
    
    \item If a team plans to constrain a resource card, it would hinder the activities, progress, and success of the other group. But only if both groups collaborate will they be able to complete the round with maximum points.
    
    \item The players are free to choose their roles in multiplayer mode. In a team, one player could hold the smartphone while the other could keep scanning the appropriate cards and exchange cards between teams if required. The other option could be one player managing the exchanges only while the other manages the rest of the play. They are free to come up with innovative task allocations of their own.
    
    \item In single-player mode, the player needs to collect as many points as possible. The highest scores are recorded and shown on a high score board, compared with recent players and all-time high scores. In multi-player mode, the team with a greater number of points wins the round.
    
\end{itemize}

\begin{figure}
  \includegraphics[width=\linewidth]{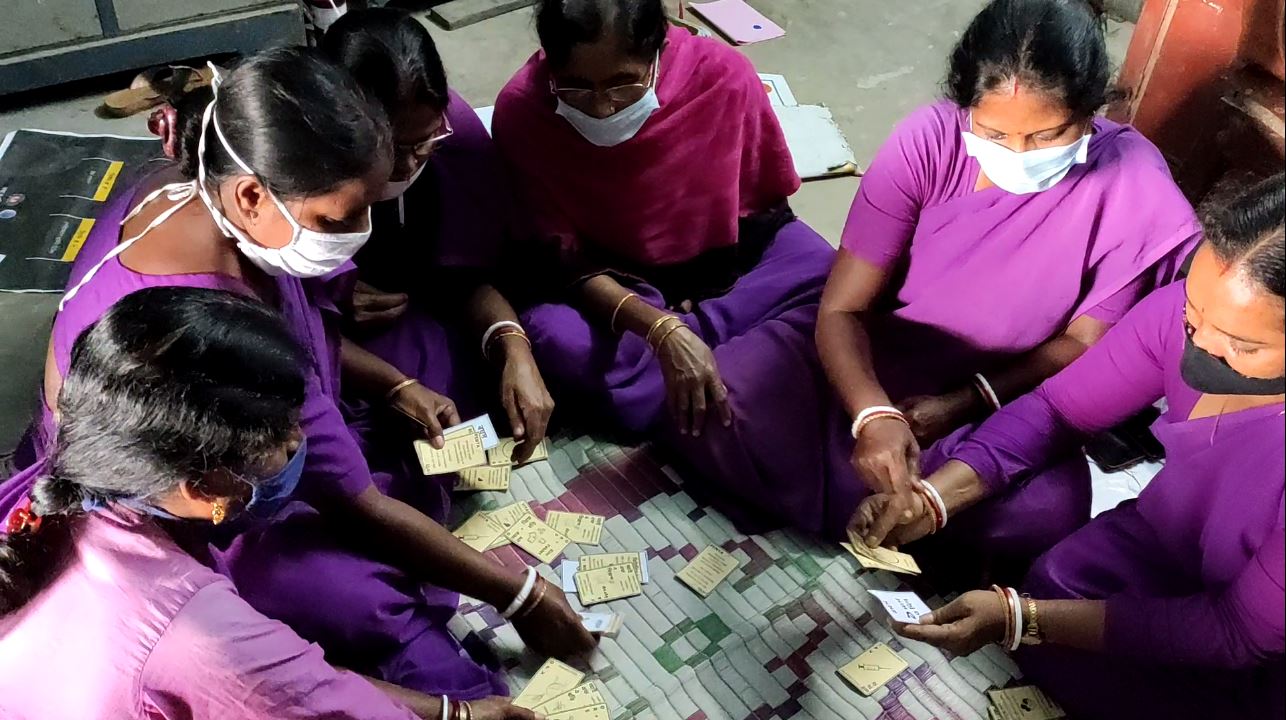}
  \caption{ASHAs play-testing with the physical cards}
  \Description{ASHAs sit in a circle forming two groups and play with a pack of physical cards}
  \label{fig:play-testing}
\end{figure}

\subsection{Correlation between physical and AR play}

The same pack of physical playing cards is used for both the physical card-playing activity and the AR-based play application. In the physical card play, the player has to read the card contents and decide to pull, put, and exchange cards to proceed through the play and win the round. In the AR-based play application, the player needs to read the card contents or the child's age, choose an age-appropriate immunization card, and place it adjacent to the child's card on the wall. In the physical card play, the playmates validate each turn of the players. In the digital application, the child avatar in AR laughs or cries indicating the right or wrong choice made by the players.

\subsection{Field Testing}

We worked in partnership with CINI, an NGO primarily working in urban areas in Kolkata and rural West Bengal. CINI works as a State Training Centre for training the ASHAs. District or sub-district-wise batches of ASHAs come and stay in CINI for about a week and get trained by State Trainers appointed by the health department under the state government. We scheduled our interventions with 5 such groups according to their training schedule. The coordinator from CINI helped in conducting the FGD and intervention. 

\subsection{Participants}
We selected 100 ASHAs through convenience sampling. 86 of them remained till the end of the study. The demographic details of the participants are given in Table \ref{tab:ParticipantsDemographics}. All participants can speak well in Bengali, their native language. Most of them could read and write in Bengali, but 12 of them could do it with difficulty.

\begin{table}[h!]
\begin{tabular}{l r}
 \toprule
 Parameters & Number of ASHA(\%) \\
 \midrule
 Age &  \\
 <30 & 4 (4.65\%) \\
 30 - 40 & 41 (47.67\%) \\
 40 - 50 & 37 (43.02\%) \\
 No information & 4 (4.65\%) \\
 \hline
 Education (grade) &  \\
 $Below  8^{th}$ & 8 (9.3\%) \\
 $8^{th}-10^{th}$ & 12 (13.95\%) \\
 $10^{th}-12^{th}$ & 56 (65.12\%) \\
 Graduation and above & 8 (9.3\%) \\
 No information & 2 (2.33\%) \\
 \hline
 Experience (years) &  \\
 0 - 5 & 4 (4.65\%) \\
 5 - 10 & 32 (37.21\%) \\
 10 - 15 & 34 (39.53\%) \\
 Above 15 & 11 (12.79\%) \\
 No information & 5 (5.81\%) \\
 
 \bottomrule
 \end{tabular}
 \caption{Participants Demographics}
 \label{tab:ParticipantsDemographics}
 \end{table}

 Participation in playtesting activity was voluntary, with an option to quit anytime during the research. All participants had their own phones from which they sent their text consent through Whatsapp or SMS to the field coordinator of the NGO, CINI. We tried to compensate the ASHAs for the valuable time and expertise they brought in. Their contribution to our research was immeasurable materialistically. As a token of gratitude, we served tea, snacks, and sweets. We checked from time to time that all ASHAs in the study followed all the COVID protocols while conducting the study.

\subsection{Evaluation}

The CHW cohort (n=86) was divided into 2 equal intervention groups (n=42 for Physical Card Intervention Group and n=44 for the AR-based App intervention group). The study conducted a paired pre-test and post-test design \citep{Dimitrov2003} to evaluate the amount of improvement that indicates the effectiveness of the created playful activities on the CHWs or ASHAs' factual and conceptual knowledge of the immunization schedule. These tests were conducted 1-2 hours before and immediately after the intervention of playing with cards and the AR app. A questionnaire containing ten questions of almost equal weightage was circulated and filled by the ASHAs. The questions tested the factual recollections of the schedule of the vaccines for children. The pre-test and post-test scores of the ASHAs were compared. The scores indicate the knowledge retention rate after conducting a particular intervention and checking it immediately after the intervention.

\subsection{Positionality}

All three authors have good exposure to conducting fieldwork with CHWs in rural India, focusing on child and maternal health. The second author has been a career bureaucrat, served the Government of India, and has been a Secretary of Women and Child Development and several other capacities in the State Government. The first and third authors are native Bengali speakers as the CHWs and belong to the same geographical region and similar socio-cultural contexts, while the third author is not a native speaker but is fluent in Bengali. We strive to understand and only partially portray the women's perspective of marginalized and low-resource contexts in the best possible manner.

\section{Findings}
\subsection{Within group and Between groups comparison}
\subsubsection{Within group comparison} After conducting the experiment with the ASHAs (n = 86), they showed an overall positive shift in post-tests scores over pretests' scores (within intervention groups) by increasing the mean score from 2.8 (SD=2.1) in the pretest to 8.6 (SD=1.2) out the total score of 10. After checking that both samples have a normal distribution of scores and no outlier, 1 sample paired t-test was conducted. $H_0$=There are no differences in existing and acquired knowledge between the intervention groups. $H_A$=There are differences in existing and acquired knowledge within groups. Null hypothesis was rejected (p<0.05,level of significance = 95\%). The analysis showed a statistically significant difference between pre and post-interventions, combining both intervention groups.

\begin{table}[h!]
  \caption{Comparison of pre-test and post-test scores combining both interventions (within group comparison). Learning through Mobile App (Intervention Group 1 or IG 1). Learning through Physical card (Intervention Group 2 or IG 2)}
  \label{tab:diff}
  \begin{tabular}{cccc}
    \toprule
    Intervention&Pre-test&Post-test&Difference\\
    \midrule
    All participants (n = 86) &2.8&8.6&5.8\\
    Standard Deviation (SD)&2.1&1.2& \\
    \bottomrule
\end{tabular}
\end{table}

\begin{table}[h!]
  \caption{Null and Alternate Hypothesis (within group comparison)}
  \label{tab:hypothesis}
  \begin{tabular}{c}
    \toprule
    \(\mu\) = Mean of Differences \\
    \midrule
    \(H_0\) = \(_\mu\) Post \(\le\) \(_\mu\) Pre \\
    \midrule
    \(H_A\) = \(_\mu\) Post > \(_\mu\) Pre \\
    \midrule
    1 sample paired t-test result = P<0.05 hence \(H_0\) rejected\\
  \bottomrule
\end{tabular}
\end{table}

\subsubsection{Between group comparison} Comparing both the intervention groups [IG1 (n=44) and IG2 (n=42)], we found that the ASHAs trained through the mobile app card play have shown an increase in the mean score from 2.7 (SD=1.1) [IG1pre] in the pretest to 9.2 (SD=2.1) [IG1post] in the post-test and the ASHAs trained through the physical card play have shown an increase in the mean score from 2.8 (SD=1.4) [IG2pre] in the pretest to 6.2 (SD=1.8) [IG2post] in the post-test out of the total score of 10. The difference of difference ( calculated by first finding differences within groups, pre-post tests, and then between two intervention groups [(IG1post - IG1pre) - (IG2post - IG2pre)] ) of mean test scores within and between groups is 3.1. After checking that both samples have a normal distribution of scores and no outlier, 2 sample t-test comparing between-group differences of difference of means within groups (pre and post-test) was conducted. $H_0$=There are no differences in knowledge gain and retention between the intervention groups. $H_A$=There are differences in knowledge retention between groups. Null hypothesis was rejected (p<0.05,level of significance = 95\%). The analysis showed a statistically significant difference between 2 intervention groups in the mean difference of pre and post-scores. This signifies that knowledge gain and immediate retention are better in AR app play intervention than in physical card play.

\begin{table}[h!]
  \caption{Summary of means of pre-test and post-test scores and their difference. Learning through Mobile App (Intervention Group 1 or IG 1). Learning through Physical Card (Intervention Group 2 or IG 2)}
  \label{tab:diff}
  \begin{tabular}{cccc}
    \toprule
    Intervention & Pre-test score & Post-test score & Difference\\
    \midrule
    IG-1 (n=44) & 2.7 (SD=1.1) & 9.2 (SD=2.1) & 6.5\\
    IG-2 (n=42) & 2.8 (SD=1.4) & 6.2 (SD=1.8) & 3.4\\
  \bottomrule
\end{tabular}
\end{table}

\begin{table}[h!]
  \caption{The difference of difference ( calculated by first finding difference within groups, pre-post tests, and then between two intervention groups [(IG1post - IG1pre) - (IG2post - IG2pre)] ) of mean test scores within and between groups}
  \label{tab:DOD}
  \begin{tabular}{cc}
    \toprule
    Difference of difference & Value\\
    \midrule
    (IG1post - IG1pre) - (IG2post - IG2pre) & 3.1\\
  \bottomrule
\end{tabular}
\end{table}

\begin{table}[h!]
  \caption{Null and Alternate Hypothesis: Comparing Knowledge Retention (KR) between pre and post-test (within groups) between 2 intervention groups (between groups)}
  \label{tab:hypothesis}
  \begin{tabular}{c}
    \toprule
    \(\mu\) = Mean of Differences \\
    \midrule
    \(H_0\) = \(_\mu\) KR-IG1 \(\le\) \(_\mu\) KR-IG2 \\
    \midrule
    \(H_A\) = \(_\mu\) KR-IG1 > \(_\mu\) KR-IG2 \\
    \midrule
    (2 samples) t-test result = P<0.05 hence \(H_0\) rejected\\
  \bottomrule
\end{tabular}
\end{table}

\subsection{Play-testing experiences}
After playing on their own smartphones 2-3 times in a loop, the players were able to memorize the immunization chart fully. They didn't require us to open the charts for reference. We followed the CHW cohort for more 6 months. We found that on reopening the app, they can quickly navigate through it. The odds of making mistakes during the play sessions reduced significantly from the fourth round of play. Through discussions with CHWs, we understood that the knowledge gain happened mostly in players who were visual learners, through reading the physical cards, and for the players who were auditory learners through listening to the other players reading aloud, which also helped in knowledge retention. After playing 10-15 rounds, the set of scenarios or challenges becomes limited to the prescribed ones similar to their training textbook, causing a reduction in fun and engagement. By adding more real-life scenarios, the play could be made more challenging. The right balance of the complexity of the scenarios and improvement of player engagement should be kept in mind while designing the experience.

\section{Novelty} None of the previous studies conducted in India have tried comparing playful experiences and the effects of physical and AR activity-based learning to train low-literate CHWs in low-resource settings. This approach of learning through the AR-based refresher training tool would make the content and immunization schedule easy to remember. The playfulness and interactive nature of the AR application would make it more exciting and engaging for the CHWs.

\section{Conclusion}

This study provides objective evidence regarding the efficacy of the refresher training tool among CHWs or the ASHAs by improving their conceptual knowledge of immunization and its schedule and strengthening their positive perception regarding the effects of immunization on child development. This research contributes to HCI by providing a new learning tool for low-literate smartphone users and empirical evidence by comparing experiment-driven activities through field observations, user testing, and interviews.

\section{Future Work}

We will apply the learning from our pilot version of the card play to our existing development. We will check whether our designed gamified interactions resulted in meaningful engagements with the CHWs and develop strategies to be incorporated to make it more effective. Also, we would like to conduct longitudinal studies assessing improvements in learning and if the knowledge acquired gets translated into their daily fieldwork in community healthcare.

\begin{acks}
We thank Science \& Engineering Research Board (SERB), Federation of Indian Chambers of Commerce \& Industry (FICCI), and UNICEF, New Delhi for supporting our research. We thank the Government of West Bengal and the Child In Need Institute (CINI) for their support in our research. We thank all the ASHAs and their supervisors who participated and supported this research. We thank the anonymous reviewers for their insightful feedback.
\end{acks}

\bibliographystyle{ACM-Reference-Format}
\bibliography{references}

\end{document}